\documentclass[aps,prd,twocolumn,showpacs]{revtex4} 

\usepackage{graphicx}

\usepackage{amsmath}
\usepackage{amssymb}

\font\FermiSmallfont=cmssq8 scaled 1200

\def\LANLppthead#1#2{
\null 
\begin{center}\vskip -1.0truein{\hbox to 7.5truein {
\hfill
\vbox to 1in {\vfill \FermiSmallfont
              \hbox{#1}
              \hbox{#2}
              \vfill}
}}\vskip-0.0truein\end{center}}

\begin{document}

\LANLppthead {LA-UR 05-9175}{astro-ph/0512631}

\title{Linear Cosmological Structure Limits on Warm
  Dark Matter}

\author{Kevork Abazajian}
\affiliation{Theoretical Division, MS B285, Los Alamos National
  Laboratory, Los Alamos, NM 87545 }

\pacs{95.35.+d,14.60.Pq,14.60.St,98.65.-r}

\begin{abstract}
  I consider constraints from observations on a cutoff scale in
  clustering due to free streaming of the dark matter in a warm dark
  matter cosmological model with a cosmological constant.  The limits
  are derived in the framework of a sterile neutrino warm dark matter
  universe, but can be applied to gravitinos and other models with
  small scale suppression in the linear matter power spectrum.  With
  freedom in all cosmological parameters including the free streaming
  scale of the sterile neutrino dark matter, limits are derived using
  observations of the fluctuations in the cosmic microwave background,
  the 3D clustering of galaxies and 1D clustering of gas in the
  Lyman-alpha (Ly$\alpha$) forest in the Sloan Digital Sky Survey
  (SDSS), as well as the Ly$\alpha$ forest in high-resolution
  spectroscopic observations.  In the most conservative case, using
  only the SDSS main-galaxy 3D power-spectrum shape, the limit is $m_s
  > 0.11\rm\ keV$; including the SDSS Ly$\alpha$ forest, this
  limit improves to $m_s > 1.7\rm\ keV$.  More stringent constraints
  may be placed from the inferred matter power spectrum from
  high-resolution Ly$\alpha$ forest observations, which has
  significant systematic uncertainties; in this case, the limit
  improves to $m_s > 3.0\rm\ keV$ (all at $95\%$ CL).
\end{abstract}

\maketitle

\section{Introduction}

The scale to which the ansatz of a (nearly) scale-invariant power
spectrum describes the cosmological primordial density fluctuations is
unquantified at the smallest scales.  The standard paradigm of an
unmodified, nearly scale-invariant power spectrum arising from the
amplification of quantum vacuum fluctuations of an inflaton field
agrees well with the inferred matter power spectrum in the linear to
mildly nonlinear regime of cosmological structure formation
\cite{Spergel:2003cb,Tegmark:2003ud,Seljak:2004xh,Viel:2004np,Abazajian:2005dt}.
Models of the nonlinear clustering of galaxies are also consistent
with a nearly scale-invariant perturbation spectrum
\cite{Abazajian:2004tn}.

There are a number of physical properties of the dark matter that may
alter its perturbation spectrum at small scales.  The hot big bang
primordial plasma in the early universe may partially or completely
thermodynamically couple to the dark matter.  Any primordial velocity
distribution that is imparted to the dark matter will allow dark
matter to escape perturbations with sufficiently small gravitational
potentials and damp structure formation at such scales.  The most
cited cold dark matter (CDM) candidate, the lightest supersymmetric
particle (LSP), has a small but non-zero velocity dispersion, which
damps structures below Earth mass scales
\cite{Hofmann:2001bi,Green:2005fa}.  In warm dark matter (WDM) models,
a light mass particle such as a sterile
neutrino~\cite{Dodelson:1993je} or gravitino~\cite{Blumenthal:1982mv}
with a large velocity dispersion can suppress linear structure
formation up to galactic scales.  The dark matter power spectrum may
also be suppressed on small scales due to the production of the dark
matter LSP through decay of charged progenitor causing a suppression
of structure below the horizon scale at
decay~\cite{Sigurdson:2003vy,Kaplinghat:2005sy} or inflation with
broken scale invariance~\cite{Kamionkowski:1999vp}. 

Models with a suppression of small scale power have drawn attention
due to their potential alleviation of several unresolved problems in
galaxy and small scale structure formation.  These include, first, the
reduction of satellite galaxy halos
\cite{Kauffmann:1993gv,Klypin:1999uc,Moore:1999wf,Willman:2004xc},
second, the reduction of galaxies in
voids~\cite{Peebles:2001nv,Bode:2000gq}, third, the low concentrations
of dark matter in
galaxies~\cite{Dalcanton:2000hn,vandenBosch:2000rz,Zentner:2002xt,Abazajian:2005kz},
fourth, the angular momentum problem of galaxy
formation~\cite{Dolgov:2001nq}, and fifth, the formation of
disk-dominated galaxies \cite{Governato:2002cv,Kormendy2005}.

A significant reduction of power on sufficiently large scales will be
in conflict with observations of clustering at small scales.  It was
noted by Narayanan et al.~\cite{Narayanan:2000tp} that observations of
power on the smallest linear scales in the Lyman-alpha (Ly$\alpha$)
forest may provide particularly stringent constraints. In this paper,
I examine the upper limits on the scale of such a suppression of small
scale power from measures of matter clustering at small scales yet in
the linear regime, where systematic effects in modeling can be
minimized.  I frame the constraints in terms of a sterile neutrino WDM
candidate, motivated by new data and an accurate calculation of the
transfer function for sterile neutrino dark matter based on an
accurate calculation of the nonthermal production of sterile neutrino
dark matter~\cite{abazajian05}.  Limits on gravitino WDM and general
departures of matter power spectra at small scales also can be
inferred from these results.  The small-scale clustering data employed
here are the Sloan Digital Sky Survey (SDSS) main galaxy 3D power
spectrum \cite{Tegmark:2003uf}, the inferred linear matter power
spectrum from observations of the flux power spectrum of Ly$\alpha$
absorption in the SDSS quasar
catalogue~\cite{McDonald:2004eu,McDonald:2004xn}, and the inferred
linear matter power spectrum from high-resolution observations of the
Ly$\alpha$ forest of Croft et al.~\cite{Croft:2000hs} as reanalyzed
and augmented by Viel et al.~\cite{Viel:2004bf} (hereafter VHS).

In order to alleviate degeneracies with other cosmological parameters,
particularly the optical depth to the cosmic microwave background
(CMB) and primordial scalar perturbation amplitude, as well as tighten
constraints within allowable possibilities, I include observations of
the fluctuations of the CMB from the first year Wilkinson Microwave
Anisotropy Probe (WMAP)~\cite{hinshaw03,kogut03}, the Cosmic
Background Imager (CBI)~\cite{readhead04},
Boomerang~\cite{Jones:2005yb}, the Arcminute Cosmology Bolometer Array
(ACBAR)~\cite{Kuo:2002ua}, and the Very Small Array
(VSA)~\cite{dickinson04}.  Viel et al.~\cite{Viel:2005qj} have done a
similar statistical analysis to that presented here, using WMAP and
the inferred linear matter power spectrum of VHS to place constraints
on a gravitino WDM candidate, which is applied to sterile neutrino
dark matter via a central-value relationship.  The work presented here
is motivated by using a significantly more accurate transfer function
for sterile neutrino dark matter of Ref.~\cite{abazajian05}, as well
as the results of inclusion of many more independent cosmological
structure data.

It should be noted that observations of nonlinear structures may be a
very powerful handle of primordial power at extremely small scales.
The observations of anomalous flux ratios in gravitational lens
systems can be an indication of substructure having a significant
fraction of the mass of the lensing galaxies, which would not be
present in certain WDM models~\cite{dalal2002,kochanek2004}.  However,
the exact level of the suppression of substructure that can be
tolerated by these lensing observations is not clear, particularly
given that some small mass halos may be formed by fragmentation in WDM
models~\cite{Knebe:2003hs}; moreover, there may be a significant
enhancement of the observed anomaly by line-of-sight isolated
halos~\cite{Chen:2003uu}.  Another highly nonlinear process asserted
to place strong constraints on the presence of small scale power is
high-redshift (high-$z$) star formation inferred to be required to
cause reionization early enough to produce the anomalously high $TE$
cross-correlation at low multipoles seen by
WMAP~\cite{Barkana2001,Yoshida:2003rm}.  CDM models themselves
generally have difficulty in producing such high-$z$ reionization even
when including the invocation of a very early generation of high-mass
star formation~\cite{Rozas:2005pt}, and it is not clear that the
resolution of the problem of an anomalously high $TE$
cross-correlation at large scales observed by WMAP will be solved by
the presence of small mass halos at high-$z$.

\section{Analysis}
\label{fluctuation}

Here I present the results of an analysis of observational constraints
on cosmological models with a cosmological constant and a general cold
to warm dark matter in the form of a sterile neutrino.  The standard
cosmological model of structure formation from adiabatic Gaussian
fluctuations seeded by an inflationary epoch is affected by
perturbation growth in the radiation through matter dominated eras.
The distribution of velocities of the dark matter suppresses
fluctuations below its free streaming scale, which increases with the
mean dark matter velocities and decreases with its mass.

The analysis here follows that of Refs.~\cite{Tegmark:2003ud} and
\cite{Seljak:2004xh}, extended to include the possibility of a sterile
neutrino warm dark matter candidate.  The strong degeneracy that is
present between the amplitude of the primordial scalar perturbations
and the optical depth to the CMB~\cite{Tegmark:2003ud} is exacerbated
by the possibility of WDM.  This results from the tightness of
constraints on the matter fluctuation amplitude arising from the
Ly$\alpha$ forest.  Allowing a reduction of power at the smallest
scales from WDM can be compensated by an increase in the primordial
fluctuation amplitude, but made consistent with the CMB fluctuation
amplitude by a high optical depth.  There results a strong correlation
between the sterile neutrino particle mass, the optical depth to the
CMB, and the primordial fluctuation amplitude.  This requires a prior
to be set on the optical depth to the CMB in order to disallow
unphysically high optical depths, $0.01<\tau<0.3$.  All other
parameters have flat priors well outside of nonzero likelihood values.

To constrain the cosmology in this analysis, I include the shape
information from the SDSS 3D power spectrum of galaxies from the SDSS
main galaxy sample~\cite{Tegmark:2003uf}, the McDonald et
al.~\cite{McDonald:2004xn} inferred linear matter power spectrum from
the analysis of the SDSS Ly$\alpha$ forest, as well as observations of
the CMB by WMAP (first year), ACBAR ($\ell>800$), CBI
($600<\ell<2000$), VSA ($\ell>600$), and BOOMERanG-2K2
\cite{hinshaw03,verde03,kogut03,Kuo:2002ua,dickinson04,readhead04,Jones:2005yb}.
I also include a reanalysis by VHS of the linear matter power spectrum
inferred from the high-resolution observations of the Ly$\alpha$
forest by the High Resolution Echelle Spectrometer of the Keck
Observatory~\cite{Croft:2000hs}.  From VHS, I also include the
inferred matter power spectrum from a new set of Ly$\alpha$ forest
data obtained with the Ultraviolet and Visual Echelle Spectrograph
(UVES) aboard the Very Large Telescope (VLT), called the Large sample
of UVES QSO Absorption Spectra (LUQAS).

\begin{figure}
\includegraphics[width=3.3truein]{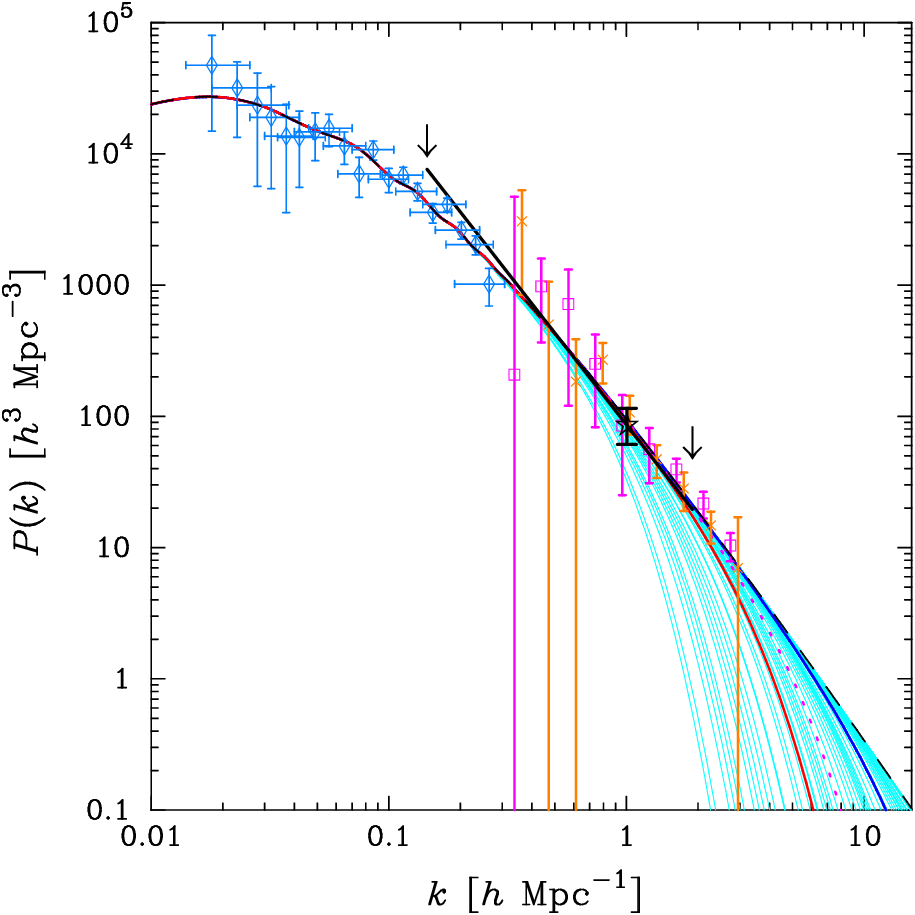}
\caption {\small Shown are the resulting linear matter power spectra
  $P(k)$ for a standard flat cosmological model $\Omega_{\rm DM} =
  0.26$, $\sigma_8 =0.9$, $\Omega_b=0.04$, and $h=0.7$ at $z=0$, and
  with sterile neutrino warm to cold dark matter in the mass range
  $0.3{\rm\ keV} < m_s < 140{\rm\ keV}$ (gray/cyan).  The
  corresponding CDM case is dashed (black).  Small-scale clustering
  data used here are the SDSS 3D power-spectrum of galaxies
  (diamonds), the inferred slope and amplitude of the matter power
  spectrum from SDSS Ly$\alpha$ forest observations (star point and
  slope between arrows), the inferred matter power spectrum from
  Ly$\alpha$ forest observations from Croft et al.~\cite{Croft:2000hs}
  (cross points) and the LUQAS (square points), as interpreted by
  VHS~\cite{Viel:2004bf}.  Ly$\alpha$ forest measures are evolved to
  $z=0$ by the appropriate growth function.  The solid (blue) line at
  high-$k$ is $P(k)$ for upper limit $m_s = 8.2\rm\ keV$ from
  observations of Virgo~\cite{Abazajian:2001vt}, the solid (red) line
  at low-$k$ is that for the lower limit from the SDSS Ly$\alpha$
  forest in this work ($m_s = 1.7\rm\ keV$), and the dotted line is
  that for the lower limit using high-resolution Ly$\alpha$ forest
  data from this work ($m_s = 3.0\rm\ keV$).
  \label{matterpower}}
\end{figure}

A useful form of the suppressed perturbation power spectrum for
sterile neutrino dark matter, $P_{\rm sterile}(k)$, relative to the
CDM case is a transfer function of the form \cite{abazajian05}
\begin{equation}
T_s(k) \equiv \sqrt{\frac{P_{\rm sterile}(k)}{P_{\rm CDM}(k)}},
\label{sterile_transfer}
\end{equation}
which can be used to convert any CDM transfer function to that of
sterile neutrino dark matter.  The fitting function that describes the
transfer function is of the form
\begin{equation}
T_s(k) = \left[1 + \left(\alpha k\right)^\nu\right]^{-\mu}, 
\label{transfer_sterile_fit}
\end{equation}
where 
\begin{equation}
\alpha = a\ \left(\frac{m_s}{1\rm\ keV}\right)^b
\left(\frac{\Omega_{\rm DM}}{0.26}\right)^c
\left(\frac{h}{0.7}\right)^d\ h^{-1}\rm\ Mpc, 
\end{equation}
and $a = 0.188$, $b=-0.858$, $c = -0.136$, $d = 0.692$, $\nu = 2.25$,
and $\mu = 3.08$.  As the sterile neutrino particle mass decreases,
the velocity distribution increases in scale, and therefore the
spatial scale $\alpha$ of the cutoff in the primordial clustering
spectrum increases (see Fig.~\ref{matterpower}).  The scale $\alpha$
represents where the power spectrum falls off by 11.8\% relative to
the standard CDM power spectrum, and can be used to place rough limits
on other nonstandard dark matter models or broken scale-invariance
inflation
models~\cite{Sigurdson:2003vy,Kaplinghat:2005sy,Kamionkowski:1999vp}.
The central values of the WDM sterile neutrino particle mass $m_s$ and
thermal gravitino particle mass $m_{\tilde g}$ can be approximately
related by~\cite{Viel:2005qj}:
\begin{equation}
m_{\tilde g} \approx 0.326\ {\rm keV}
\left(\frac{m_s}{1\rm\ keV}\right)^{3/4}
\left(\frac{\Omega_{\rm DM}h^2}{0.12}\right)^{1/4}.
\end{equation}

\begin{figure}
\includegraphics[width=3.3truein]{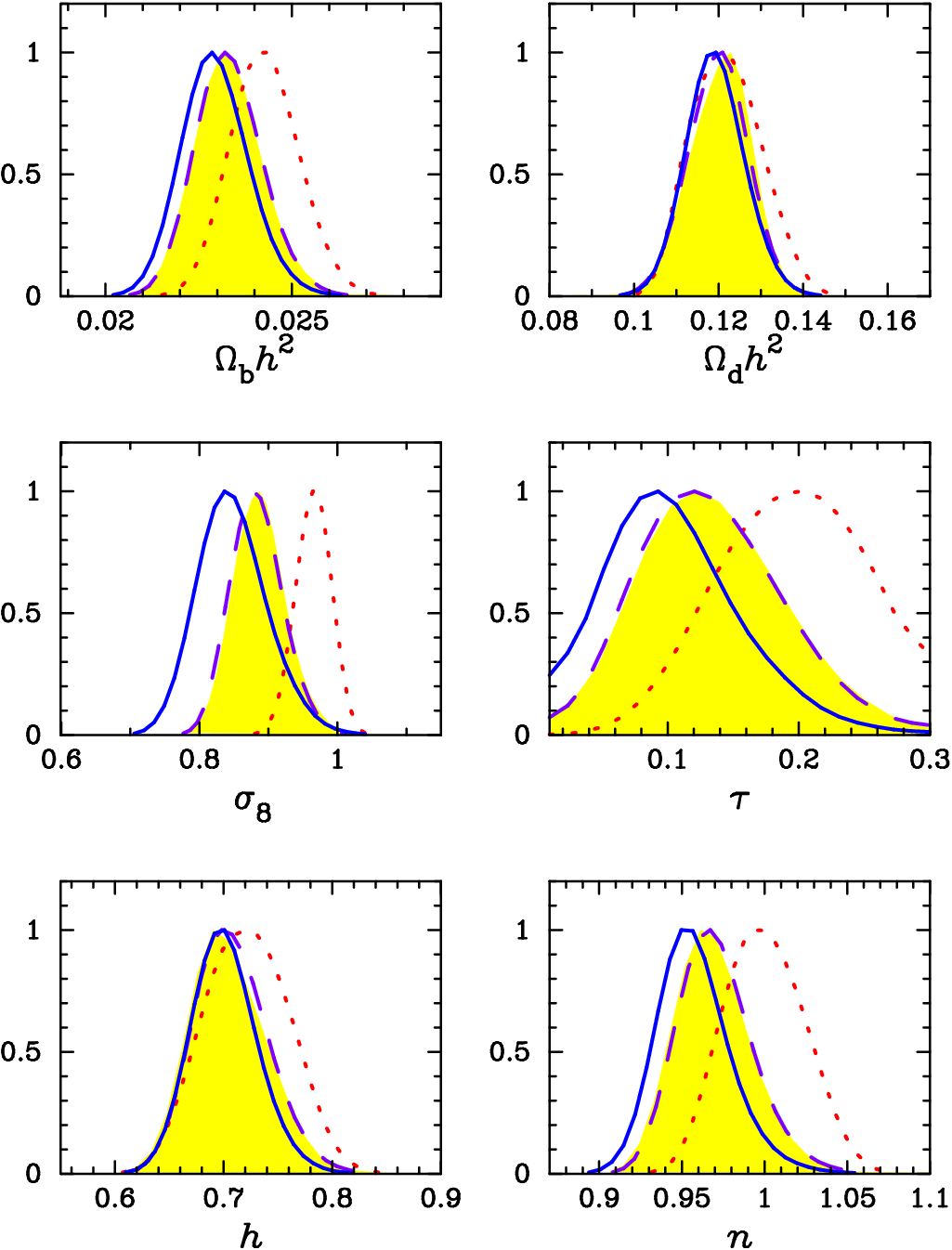}
\caption
{\small Shown here are the 1D marginalized likelihoods on chosen
  base and derived cosmological parameters for the CMB plus SDSS 3D
  $P_g(k)$ in solid (blue), plus SDSS Ly$\alpha$ forest in dashed
  (purple), plus the VHS high-resolution Ly$\alpha$ forest in dotted
  (red).  The shaded (yellow) likelihoods are for the CMB plus SDSS 3D
  $P_g(k)$ plus SDSS Ly$\alpha$ forest for the standard CDM case,
  which shows that parameters are generally not biased when including
  the presence of WDM.
\label{cosmo_like}}
\end{figure}

For a given cosmological parameter and sterile neutrino particle mass choice,
we use the predicted matter power spectrum and CMB anisotropy angular
correlation functions to calculate the likelihood to observe the
combined data.  The parameters we vary are the six parameters for the
``vanilla'' standard cosmological model plus the inverse sterile
neutrino mass: 
\begin{equation}
\mathbf p = (\Omega_b h^2, \Omega_d
h^2,\Theta_s,\ln(A),n,\tau,m_s^{-1}),
\label{params}
\end{equation}
where $\Omega_b$ and $\Omega_d$ are fractions of the critical density
in baryons and dark matter; $\Theta_s$ is the angular acoustic peak
scale of the CMB, a useful proxy for the Hubble parameter, $H_0 =
100\, h \rm\ km\;s^{-1}\,Mpc^{-1}$; $A$ and $n$ are the amplitude and
tilt of the primordial scalar fluctuations; $\tau$ is the optical
depth due to reionization.  The inverse of $m_s$ is used to allow for
a flat prior on and sampling of arbitrarily massive sterile neutrino
models. To measure the likelihood space allowed by the data, we use a
Markov Chain Monte Carlo (MCMC) method with a modified version of the
\citet{lewis02} CosmoMC code.  We use the WMAP team's code to
calculate the WMAP observations' likelihood, and CosmoMC to calculate
that for Boomerang, ACBAR, CBI, and VSA. After burn-in, the chains
typically sample $\gtrsim 10^5$ points, and convergence and likelihood
statistics are calculated from these.  Resulting marginalized 1D
likelihoods for different data sets are shown in
Fig.~\ref{cosmo_like}.

\begin{figure}
\includegraphics[width=3.3truein]{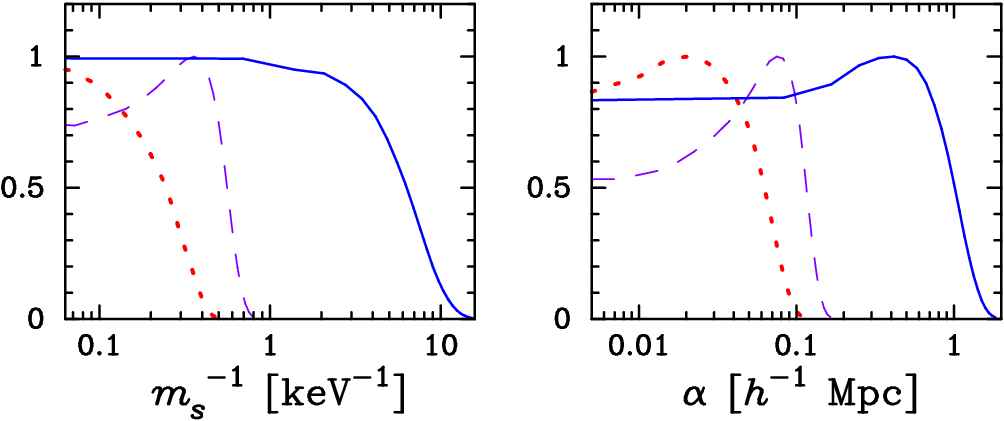}
\caption
{\small Shown here are the 1D marginalized likelihoods for the
  (inverse) mass of the sterile neutrino dark matter, $m_s$ and the
  scale of the primordial spectrum cutoff $\alpha$, from the CMB plus
  SDSS 3D $P_g(k)$ in solid (blue), plus SDSS Ly$\alpha$ forest in
  dashed (purple), and plus the VHS high-resolution Ly$\alpha$ forest
  in dotted (red).
\label{like_ms}}
\end{figure}

The shape of the 3D galaxy power spectrum from the main galaxy
catalogue of the SDSS, $P_g(k)$ is a robust measure of the shape of
the 3D matter power spectrum in the linear regime.  The process of
galaxy formation occurs on scales $\ll 10\rm\ Mpc$~\cite{White1978},
where gas cools within dark matter halos and the large scale
clustering then follows that of the total matter
components~\cite{Weinberg:2002rm}.  Therefore, using the shape
information from the 3D power spectrum of SDSS galaxies at $k
< 0.2\ h^{-1}\rm Mpc$ is a very conservative upper bound on the
deviations of the primordial spectrum due to WDM.  The lower limit on
a sterile neutrino dark matter candidate in this case is $m_s >
0.108\rm\ keV$, gravitino WDM particle mass $m_{\tilde g} >
0.063\rm\ keV$ and cutoff scale $\alpha < 1.22\ h^{-1}\rm Mpc$, at
$95\%\ \rm CL$, with 1D likelihoods shown in Fig.~\ref{like_ms}.

Observations of spectra of distant quasars contain absorption features
from the Ly$\alpha$ transition due to neutral hydrogen gas along the
line of site.  The numerous absorption features are a Ly$\alpha$
``forest'' that follows gas in-fall into gravitational potentials along
the line of site and therefore can be used to infer the power spectrum
of dark matter density along this line of site.  The statistic chosen
to parametrize the clustering of the Ly$\alpha$ features is the power
spectrum of the transmitted flux fraction $P_F(k,z)$.  The relation
between the gas and dark matter density relies on the state of
ionization equilibrium of the gas, and in turn is dependent on the
temperature-density relation, the intensity of the UV background, the
mean baryon density and many other physical parameters.

With tests via hydrodynamic simulations, one approach is mapping the
relation between the 1D $P_F(k,z)$ and 3D matter $P(k,z)$, which then
can be inverted through a biasing relation $P_F(k,z)= b^2(k,z)
P(k,z)$.  This is the approach of VHS, is what is used by Viel et
al.~\cite{Viel:2005qj} to place a limit on the WDM free streaming
scale and lower limit on the WDM particle mass, and is my method of
applying their results here.  A key result of Ref.~\cite{Viel:2005qj}
is that $b^2(k,z)$ is nearly identical for CDM and WDM over the
wave-numbers used in the inversion of $P_F(k,z)$.  This motivates
applying the results of CDM-based analyses to a WDM model, though the
differences between WDM and CDM models in power spectrum inversion
should be explored further.  It is important to note that there is
much contention regarding the accuracy of the inversion $b^2(k,z)$
(see, e.g.,
Refs.~\cite{Zaldarriaga:2001xs,Seljak:2003jg,Jena:2004fc}), especially
regarding mapping the full parameter space of the inversion $b^2(k,
{\mathbf p, \mathbf q})$ in cosmological $\mathbf p$ and physical
parameters $\mathbf q$ of the gas.

Instead of a direct relation between the flux and matter power
spectra, McDonald et al.~\cite{McDonald:2004xn} map out the likelihood
space for the amplitude and slope of the linear matter power spectrum
for the SDSS Ly$\alpha$ forest data for a given range of physical
parameters, in order to quantify much of the systematic effects of the
uncertain physical states of the gas in the modeling.  Note that Viel
et al.~\cite{Viel:2005ha} have done a reanalysis of the inferred
matter power spectrum using a very different analysis and get
consistent results with McDonald et al.  Since the systematic effects
are generally non-Gaussian in likelihood space, McDonald et
al. therefore provide a likelihood look-up table for the amplitude and
slope of the matter power spectrum at the effective redshift of the
SDSS Ly$\alpha$ forest.  Using this likelihood from this
interpretation of the SDSS Ly$\alpha$ forest data, combined with the
base model of the CMB plus SDSS 3D $P_g(k)$, provides an order of
magnitude more stringent constraint on the free streaming scale and
WDM particle mass: $m_s > 1.71\rm\ keV$, $m_{\tilde g} > 0.500\rm\
keV$ and $\alpha < 0.121\ h^{-1}\rm Mpc$, at $95\%\ \rm CL$.  It
should be noted that the 1D likelihood for the cutoff scale and
inverse sterile neutrino mass peak at non-vanishing values
(Fig.~\ref{like_ms}), and indicate the possibility of detecting a
small scale cutoff through Ly$\alpha$ forest observations.

\begin{figure}
\includegraphics[width=3.3truein]{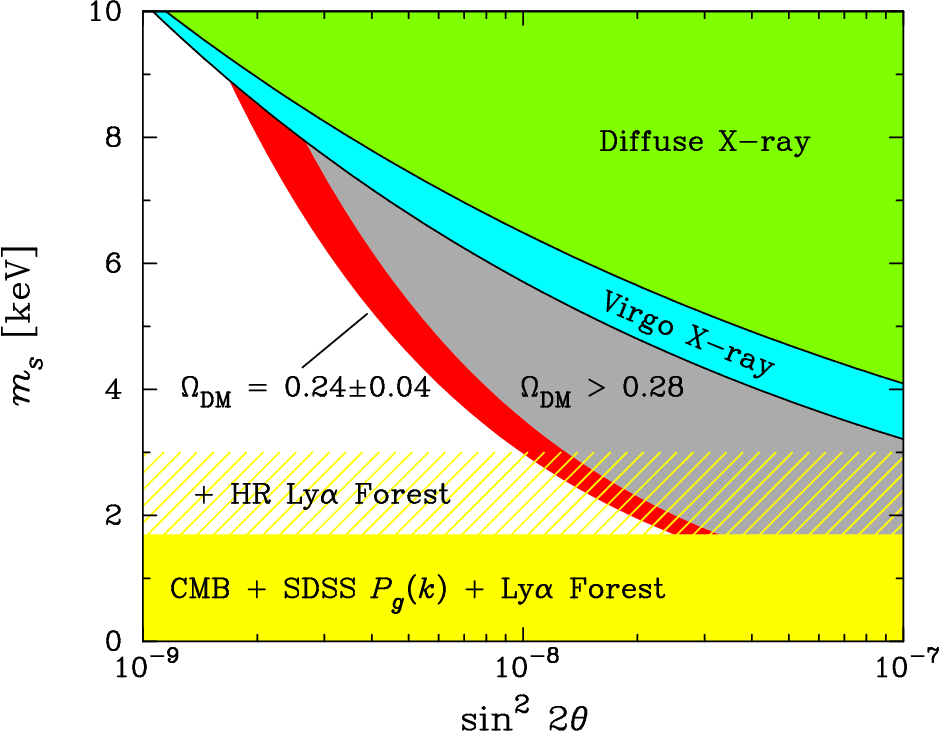}
\caption {\small The parameter space for sterile neutrino dark matter
  production and constraints.  The red (dark grey) band is consistent
  with sterile neutrinos composing the dark matter consistent with
  $\Omega_{\rm DM} = 0.24\pm 0.04$~\cite{abazajian05}.  Upper limits
  arise from radiative decay limits from XMM-Newton X-ray observations 
  of the Virgo
  Cluster~\cite{Abazajian:2001vt} and the diffuse X-ray 
  background~\cite{Boyarsky:2005us}.  The lower limits of this work
  from the CMB plus SDSS 3D $P_g(k)$ plus SDSS Ly$\alpha$ are in solid
  (yellow) at $m_s<1.7\rm\ keV$ and all previous data plus VHS in
  dashed (yellow) at $m_s<3.0\rm\ keV$.
  \label{omega_bands}}
\end{figure}

The high-resolution Ly$\alpha$ forest data analyzed by VHS can provide
an even more stringent constraint on the free streaming scale of the
WDM and particle mass, and I include that data as done in
Ref.~\cite{Viel:2005qj}.  I use the inferred matter power spectrum of
VHS, Table 4, but reduced by a uniform 7\% for a change in the
temperature-density relation, as done in
Refs.~\cite{Viel:2004np,Viel:2005qj}.  In addition, there are
systematic uncertainties as described in
Refs.~\cite{Viel:2004np,Viel:2005qj}, namely for the power and
amplitude of the temperature density relation (5\%), the measure of
the effective optical depth for calibration (8\%), the uncertainty in
the $b^2(k,z)$ inversion (5\%), uncertainty in the effects of galactic
winds (5\%), and finally an uncertainty due to differences in
numerical simulations (8\%).  These systematic errors are added in
quadrature to give an overall systematic uncertainty of $14.5\%$,
which is comparable to the statistical uncertainties to which it is,
in turn, added in quadrature.  It is certainly unknown how correct the
assumption is that these systematic errors are Gaussian and should be
combined in quadrature.  However, a more robust analysis of the
high-resolution Ly$\alpha$ forest data does not yet exist, and so I
follow the procedure of using the VHS data as in
Refs.~\cite{Viel:2004np,Viel:2005qj}, with the above caveats in mind.
Using the VHS high-resolution data in concert with the SDSS 3D
$P_g(k)$ and SDSS Ly$\alpha$ forest gives the constraints: $m_s >
3.00\rm\ keV$, $m_{\tilde g} > 0.769\rm\ keV$ and $\alpha <
0.0756\ h^{-1}\rm Mpc$, at $95\%\ \rm CL$.  A summary of the parameter
space for sterile neutrino dark matter is shown in Fig.~\ref{omega_bands}.

\section{Conclusions}
\label{conclusions}

Potential problems in galaxy and small-scale structure formation
indicate the possibility of a small-scale velocity damping of
perturbations of the type in warm dark matter such a sterile neutrino
dark matter.  This work presents a calculation of cosmological
parameter constraints including the possibility of a warm dark matter
candidate to investigate the largest scale of the suppression of small
scale power allowed by linear measures of the cosmological structure
from the inferred matter power spectrum.  I find the lower limits on a
sterile neutrino dark matter particle mass placed by observations of
fluctuations in the cosmic microwave background, clustering of
galaxies and in the Ly$\alpha$ forest in the Sloan Digital Sky Survey
(SDSS), and clustering of the Ly$\alpha$ forest in high-resolution
spectroscopic observations by the Keck and VLT Observatories.  The
lower mass bound in the most conservative case, using only the SDSS
main-galaxy 3D power-spectrum shape is $m_s > 0.11\rm\ keV$; including
the SDSS Ly$\alpha$ forest, this limit improves to $m_s > 1.7\rm\
keV$, while also including high-resolution Ly$\alpha$ forest data,
this improves to $m_s > 3.0\rm\ keV$, with all limits at $95\%$ CL.
These results are consistent with the results of Viel et
al.~\cite{Viel:2005qj}, but more constrained than their results ($m_s
> 2.0\rm\ keV$) due to the inclusion here of additional CMB and large
scale structure data. The limits from the high-resolution Ly$\alpha$
forest data are subject to large systematic uncertainties, and
therefore I adopt the SDSS Ly$\alpha$ data as the current WDM
structure benchmark.  These constraints can also be inferred to place
limits on gravitino warm dark matter candidates as well as other
models with a primordial fluctuation small scale cutoff, $\alpha$.
The central values and confidence intervals for the other cosmological
parameters do not change significantly with the inclusion of the
possibility of WDM.

The sterile neutrino WDM candidate is well constrained by the linear
clustering observations considered here, and by limits on its
radiative decay by observations of the Virgo cluster by XMM-Newton,
which provides an upper mass
constraint~\cite{Abazajian:2001vt,Boehringer2001}.  There also exist
upper mass constraints from the diffuse X-ray
background~\cite{Mapelli:2005hq,Boyarsky:2005us}.  The combination of these
constraints allows a narrow range for this particle dark matter
candidate mass:
\begin{equation}
1.7{\rm\ keV} < m_s < 8.2\rm\ keV.
\end{equation}
It remains for further work to detect or constrain X-ray lines from
nearby clusters of galaxies, as well as use the presence or lack of a
cutoff scale in the linear matter power spectrum to detect or exclude
this dark matter candidate.

\acknowledgments

It is a pleasure to thank Neal Dalal, Salman Habib, Katrin Heitmann,
Julien Lesgourgues and Matteo Viel for useful discussions.  This work
was supported by Los Alamos National Laboratory under DOE contract
W-7405-ENG-36.

{\it Note added in proof:} A recent preprint, Seljak et
al. astro-ph/0602430, has suggested significantly more stringent
constraints on the free streaming scale of the WDM than that presented
here ($m_s > 14\rm\ keV,\ 95\%\ CL$).  The essential difference in that
work is claimed to be the higher spatial resolution in their numerical
simulations which resolve the high sensitivity of the 1D flux power
spectrum to a reduction in small-scale power relative to the 3D power
spectrum.  However, the spatial resolution of the grid in simulations
in Narayanan et al [27] is higher ($49\ h^{-1}\rm kpc$) than that in
astro-ph/0602430 ($78\ h^{-1}\rm kpc$), yet Narayanan et al. see
essentially no difference in high-resolution flux power spectra
between CDM and $1\rm\ keV$ thermal WDM ($4.4\rm\ keV$ sterile
neutrinos).  Therefore, it is not clear from where the claimed
stringent constraints in astro-ph/0602430 arise.

\bibliography{sndm}

\begin{thebibliography}{59}
\expandafter\ifx\csname natexlab\endcsname\relax\def\natexlab#1{#1}\fi
\expandafter\ifx\csname bibnamefont\endcsname\relax
  \def\bibnamefont#1{#1}\fi
\expandafter\ifx\csname bibfnamefont\endcsname\relax
  \def\bibfnamefont#1{#1}\fi
\expandafter\ifx\csname citenamefont\endcsname\relax
  \def\citenamefont#1{#1}\fi
\expandafter\ifx\csname url\endcsname\relax
  \def\url#1{\texttt{#1}}\fi
\expandafter\ifx\csname urlprefix\endcsname\relax\def\urlprefix{URL }\fi
\providecommand{\bibinfo}[2]{#2}
\providecommand{\eprint}[2][]{\url{#2}}

\bibitem[{\citenamefont{Spergel et~al.}(2003)}]{Spergel:2003cb}
\bibinfo{author}{\bibfnamefont{D.~N.} \bibnamefont{Spergel}}
  \bibnamefont{et~al.} (\bibinfo{collaboration}{WMAP}),
  \bibinfo{journal}{Astrophys. J. Suppl.} \textbf{\bibinfo{volume}{148}},
  \bibinfo{pages}{175} (\bibinfo{year}{2003}), \eprint{astro-ph/0302209}.

\bibitem[{\citenamefont{Tegmark et~al.}(2004{\natexlab{a}})}]{Tegmark:2003ud}
\bibinfo{author}{\bibfnamefont{M.}~\bibnamefont{Tegmark}} \bibnamefont{et~al.}
  (\bibinfo{collaboration}{SDSS}), \bibinfo{journal}{Phys. Rev.}
  \textbf{\bibinfo{volume}{D69}}, \bibinfo{pages}{103501}
  (\bibinfo{year}{2004}{\natexlab{a}}), \eprint{astro-ph/0310723}.

\bibitem[{\citenamefont{Seljak et~al.}(2005)}]{Seljak:2004xh}
\bibinfo{author}{\bibfnamefont{U.}~\bibnamefont{Seljak}} \bibnamefont{et~al.},
  \bibinfo{journal}{Phys. Rev.} \textbf{\bibinfo{volume}{D71}},
  \bibinfo{pages}{103515} (\bibinfo{year}{2005}), \eprint{astro-ph/0407372}.

\bibitem[{\citenamefont{Viel et~al.}(2004{\natexlab{a}})\citenamefont{Viel,
  Weller, and Haehnelt}}]{Viel:2004np}
\bibinfo{author}{\bibfnamefont{M.}~\bibnamefont{Viel}},
  \bibinfo{author}{\bibfnamefont{J.}~\bibnamefont{Weller}}, \bibnamefont{and}
  \bibinfo{author}{\bibfnamefont{M.}~\bibnamefont{Haehnelt}},
  \bibinfo{journal}{Mon. Not. Roy. Astron. Soc.}
  \textbf{\bibinfo{volume}{355}}, \bibinfo{pages}{L23}
  (\bibinfo{year}{2004}{\natexlab{a}}), \eprint{astro-ph/0407294}.

\bibitem[{\citenamefont{Abazajian
  et~al.}(2005{\natexlab{a}})\citenamefont{Abazajian, Kadota, and
  Stewart}}]{Abazajian:2005dt}
\bibinfo{author}{\bibfnamefont{K.}~\bibnamefont{Abazajian}},
  \bibinfo{author}{\bibfnamefont{K.}~\bibnamefont{Kadota}}, \bibnamefont{and}
  \bibinfo{author}{\bibfnamefont{E.~D.} \bibnamefont{Stewart}},
  \bibinfo{journal}{JCAP} \textbf{\bibinfo{volume}{0508}}, \bibinfo{pages}{008}
  (\bibinfo{year}{2005}{\natexlab{a}}), \eprint{astro-ph/0507224}.

\bibitem[{\citenamefont{Abazajian
  et~al.}(2005{\natexlab{b}})}]{Abazajian:2004tn}
\bibinfo{author}{\bibfnamefont{K.}~\bibnamefont{Abazajian}}
  \bibnamefont{et~al.}, \bibinfo{journal}{Astrophys. J.}
  \textbf{\bibinfo{volume}{625}}, \bibinfo{pages}{613}
  (\bibinfo{year}{2005}{\natexlab{b}}), \eprint{astro-ph/0408003}.

\bibitem[{\citenamefont{Hofmann et~al.}(2001)\citenamefont{Hofmann, Schwarz,
  and Stoecker}}]{Hofmann:2001bi}
\bibinfo{author}{\bibfnamefont{S.}~\bibnamefont{Hofmann}},
  \bibinfo{author}{\bibfnamefont{D.~J.} \bibnamefont{Schwarz}},
  \bibnamefont{and} \bibinfo{author}{\bibfnamefont{H.}~\bibnamefont{Stoecker}},
  \bibinfo{journal}{Phys. Rev.} \textbf{\bibinfo{volume}{D64}},
  \bibinfo{pages}{083507} (\bibinfo{year}{2001}), \eprint{astro-ph/0104173}.

\bibitem[{\citenamefont{Green et~al.}(2005)\citenamefont{Green, Hofmann, and
  Schwarz}}]{Green:2005fa}
\bibinfo{author}{\bibfnamefont{A.~M.} \bibnamefont{Green}},
  \bibinfo{author}{\bibfnamefont{S.}~\bibnamefont{Hofmann}}, \bibnamefont{and}
  \bibinfo{author}{\bibfnamefont{D.~J.} \bibnamefont{Schwarz}},
  \bibinfo{journal}{JCAP} \textbf{\bibinfo{volume}{0508}}, \bibinfo{pages}{003}
  (\bibinfo{year}{2005}), \eprint{astro-ph/0503387}.

\bibitem[{\citenamefont{Dodelson and Widrow}(1994)}]{Dodelson:1993je}
\bibinfo{author}{\bibfnamefont{S.}~\bibnamefont{Dodelson}} \bibnamefont{and}
  \bibinfo{author}{\bibfnamefont{L.~M.} \bibnamefont{Widrow}},
  \bibinfo{journal}{Phys. Rev. Lett.} \textbf{\bibinfo{volume}{72}},
  \bibinfo{pages}{17} (\bibinfo{year}{1994}), \eprint{hep-ph/9303287}.

\bibitem[{\citenamefont{Blumenthal et~al.}(1982)\citenamefont{Blumenthal,
  Pagels, and Primack}}]{Blumenthal:1982mv}
\bibinfo{author}{\bibfnamefont{G.~R.} \bibnamefont{Blumenthal}},
  \bibinfo{author}{\bibfnamefont{H.}~\bibnamefont{Pagels}}, \bibnamefont{and}
  \bibinfo{author}{\bibfnamefont{J.~R.} \bibnamefont{Primack}},
  \bibinfo{journal}{Nature} \textbf{\bibinfo{volume}{299}}, \bibinfo{pages}{37}
  (\bibinfo{year}{1982}).

\bibitem[{\citenamefont{Sigurdson and Kamionkowski}(2004)}]{Sigurdson:2003vy}
\bibinfo{author}{\bibfnamefont{K.}~\bibnamefont{Sigurdson}} \bibnamefont{and}
  \bibinfo{author}{\bibfnamefont{M.}~\bibnamefont{Kamionkowski}},
  \bibinfo{journal}{Phys. Rev. Lett.} \textbf{\bibinfo{volume}{92}},
  \bibinfo{pages}{171302} (\bibinfo{year}{2004}), \eprint{astro-ph/0311486}.

\bibitem[{\citenamefont{Kaplinghat}(2005)}]{Kaplinghat:2005sy}
\bibinfo{author}{\bibfnamefont{M.}~\bibnamefont{Kaplinghat}},
  \bibinfo{journal}{Phys. Rev.} \textbf{\bibinfo{volume}{D72}},
  \bibinfo{pages}{063510} (\bibinfo{year}{2005}), \eprint{astro-ph/0507300}.

\bibitem[{\citenamefont{Kamionkowski and Liddle}(2000)}]{Kamionkowski:1999vp}
\bibinfo{author}{\bibfnamefont{M.}~\bibnamefont{Kamionkowski}}
  \bibnamefont{and} \bibinfo{author}{\bibfnamefont{A.~R.}
  \bibnamefont{Liddle}}, \bibinfo{journal}{Phys. Rev. Lett.}
  \textbf{\bibinfo{volume}{84}}, \bibinfo{pages}{4525} (\bibinfo{year}{2000}),
  \eprint{astro-ph/9911103}.

\bibitem[{\citenamefont{Kauffmann et~al.}(1993)\citenamefont{Kauffmann, White,
  and Guiderdoni}}]{Kauffmann:1993gv}
\bibinfo{author}{\bibfnamefont{G.}~\bibnamefont{Kauffmann}},
  \bibinfo{author}{\bibfnamefont{S.~D.~M.} \bibnamefont{White}},
  \bibnamefont{and}
  \bibinfo{author}{\bibfnamefont{B.}~\bibnamefont{Guiderdoni}},
  \bibinfo{journal}{Mon. Not. Roy. Astron. Soc.}
  \textbf{\bibinfo{volume}{264}}, \bibinfo{pages}{201} (\bibinfo{year}{1993}).

\bibitem[{\citenamefont{Klypin et~al.}(1999)\citenamefont{Klypin, Kravtsov,
  Valenzuela, and Prada}}]{Klypin:1999uc}
\bibinfo{author}{\bibfnamefont{A.~A.} \bibnamefont{Klypin}},
  \bibinfo{author}{\bibfnamefont{A.~V.} \bibnamefont{Kravtsov}},
  \bibinfo{author}{\bibfnamefont{O.}~\bibnamefont{Valenzuela}},
  \bibnamefont{and} \bibinfo{author}{\bibfnamefont{F.}~\bibnamefont{Prada}},
  \bibinfo{journal}{Astrophys. J.} \textbf{\bibinfo{volume}{522}},
  \bibinfo{pages}{82} (\bibinfo{year}{1999}), \eprint{astro-ph/9901240}.

\bibitem[{\citenamefont{Moore et~al.}(1999)}]{Moore:1999wf}
\bibinfo{author}{\bibfnamefont{B.}~\bibnamefont{Moore}} \bibnamefont{et~al.},
  \bibinfo{journal}{Astrophys. J.} \textbf{\bibinfo{volume}{524}},
  \bibinfo{pages}{L19} (\bibinfo{year}{1999}), \eprint{astro-ph/9907411}.

\bibitem[{\citenamefont{Willman et~al.}(2004)\citenamefont{Willman, Governato,
  Wadsley, and Quinn}}]{Willman:2004xc}
\bibinfo{author}{\bibfnamefont{B.}~\bibnamefont{Willman}},
  \bibinfo{author}{\bibfnamefont{F.}~\bibnamefont{Governato}},
  \bibinfo{author}{\bibfnamefont{J.}~\bibnamefont{Wadsley}}, \bibnamefont{and}
  \bibinfo{author}{\bibfnamefont{T.}~\bibnamefont{Quinn}},
  \bibinfo{journal}{Mon. Not. Roy. Astron. Soc.}
  \textbf{\bibinfo{volume}{353}}, \bibinfo{pages}{639} (\bibinfo{year}{2004}),
  \eprint{astro-ph/0403001}.

\bibitem[{\citenamefont{Peebles}(2001)}]{Peebles:2001nv}
\bibinfo{author}{\bibfnamefont{P.~J.~E.} \bibnamefont{Peebles}}
  (\bibinfo{year}{2001}), \eprint{astro-ph/0101127}.

\bibitem[{\citenamefont{Bode et~al.}(2001)\citenamefont{Bode, Ostriker, and
  Turok}}]{Bode:2000gq}
\bibinfo{author}{\bibfnamefont{P.}~\bibnamefont{Bode}},
  \bibinfo{author}{\bibfnamefont{J.~P.} \bibnamefont{Ostriker}},
  \bibnamefont{and} \bibinfo{author}{\bibfnamefont{N.}~\bibnamefont{Turok}},
  \bibinfo{journal}{Astrophys. J.} \textbf{\bibinfo{volume}{556}},
  \bibinfo{pages}{93} (\bibinfo{year}{2001}), \eprint{astro-ph/0010389}.

\bibitem[{\citenamefont{Dalcanton and Hogan}(2001)}]{Dalcanton:2000hn}
\bibinfo{author}{\bibfnamefont{J.~J.} \bibnamefont{Dalcanton}}
  \bibnamefont{and} \bibinfo{author}{\bibfnamefont{C.~J.} \bibnamefont{Hogan}},
  \bibinfo{journal}{Astrophys. J.} \textbf{\bibinfo{volume}{561}},
  \bibinfo{pages}{35} (\bibinfo{year}{2001}), \eprint{astro-ph/0004381}.

\bibitem[{\citenamefont{van~den Bosch and Swaters}(2001)}]{vandenBosch:2000rz}
\bibinfo{author}{\bibfnamefont{F.~C.} \bibnamefont{van~den Bosch}}
  \bibnamefont{and} \bibinfo{author}{\bibfnamefont{R.~A.}
  \bibnamefont{Swaters}}, \bibinfo{journal}{Mon. Not. Roy. Astron. Soc.}
  \textbf{\bibinfo{volume}{325}}, \bibinfo{pages}{1017} (\bibinfo{year}{2001}),
  \eprint{astro-ph/0006048}.

\bibitem[{\citenamefont{Zentner and Bullock}(2002)}]{Zentner:2002xt}
\bibinfo{author}{\bibfnamefont{A.~R.} \bibnamefont{Zentner}} \bibnamefont{and}
  \bibinfo{author}{\bibfnamefont{J.~S.} \bibnamefont{Bullock}},
  \bibinfo{journal}{Phys. Rev.} \textbf{\bibinfo{volume}{D66}},
  \bibinfo{pages}{043003} (\bibinfo{year}{2002}), \eprint{astro-ph/0205216}.

\bibitem[{\citenamefont{Abazajian
  et~al.}(2005{\natexlab{c}})\citenamefont{Abazajian, Koushiappas, and
  Zentner}}]{Abazajian:2005kz}
\bibinfo{author}{\bibfnamefont{K.}~\bibnamefont{Abazajian}},
  \bibinfo{author}{\bibfnamefont{S.~M.} \bibnamefont{Koushiappas}},
  \bibnamefont{and} \bibinfo{author}{\bibfnamefont{A.~R.}
  \bibnamefont{Zentner}}, \bibinfo{journal}{in preparation}
  (\bibinfo{year}{2005}{\natexlab{c}}).

\bibitem[{\citenamefont{Dolgov and Sommer-Larsen}(2001)}]{Dolgov:2001nq}
\bibinfo{author}{\bibfnamefont{A.~D.} \bibnamefont{Dolgov}} \bibnamefont{and}
  \bibinfo{author}{\bibfnamefont{J.}~\bibnamefont{Sommer-Larsen}},
  \bibinfo{journal}{Astrophys. J.} \textbf{\bibinfo{volume}{551}},
  \bibinfo{pages}{608} (\bibinfo{year}{2001}).

\bibitem[{\citenamefont{Governato et~al.}(2004)}]{Governato:2002cv}
\bibinfo{author}{\bibfnamefont{F.}~\bibnamefont{Governato}}
  \bibnamefont{et~al.}, \bibinfo{journal}{Astrophys. J.}
  \textbf{\bibinfo{volume}{607}}, \bibinfo{pages}{688} (\bibinfo{year}{2004}),
  \eprint{astro-ph/0207044}.

\bibitem[{\citenamefont{{Kormendy} and {Fisher}}(2005)}]{Kormendy2005}
\bibinfo{author}{\bibfnamefont{J.}~\bibnamefont{{Kormendy}}} \bibnamefont{and}
  \bibinfo{author}{\bibfnamefont{D.~B.} \bibnamefont{{Fisher}}},
  \bibinfo{journal}{Revista Mexicana de Astronomia y Astrofisica (Conference
  Series)} \textbf{\bibinfo{volume}{25}}, \bibinfo{pages}{101}
  (\bibinfo{year}{2005}), \eprint{astro-ph/0507525}.

\bibitem[{\citenamefont{Narayanan et~al.}(2000)\citenamefont{Narayanan,
  Spergel, Dave, and Ma}}]{Narayanan:2000tp}
\bibinfo{author}{\bibfnamefont{V.~K.} \bibnamefont{Narayanan}},
  \bibinfo{author}{\bibfnamefont{D.~N.} \bibnamefont{Spergel}},
  \bibinfo{author}{\bibfnamefont{R.}~\bibnamefont{Dave}}, \bibnamefont{and}
  \bibinfo{author}{\bibfnamefont{C.-P.} \bibnamefont{Ma}},
  \bibinfo{journal}{Astrophys. J.} \textbf{\bibinfo{volume}{543}},
  \bibinfo{pages}{L103} (\bibinfo{year}{2000}), \eprint{astro-ph/0005095}.

\bibitem[{\citenamefont{Abazajian}(2005)}]{abazajian05}
\bibinfo{author}{\bibfnamefont{K.}~\bibnamefont{Abazajian}},
  \bibinfo{journal}{[Phys.\ Rev.\ D (to be published)]}
  (\bibinfo{year}{2005}), \eprint{astro-ph/0511630}.

\bibitem[{\citenamefont{Tegmark et~al.}(2004{\natexlab{b}})}]{Tegmark:2003uf}
\bibinfo{author}{\bibfnamefont{M.}~\bibnamefont{Tegmark}} \bibnamefont{et~al.}
  (\bibinfo{collaboration}{SDSS}), \bibinfo{journal}{Astrophys. J.}
  \textbf{\bibinfo{volume}{606}}, \bibinfo{pages}{702}
  (\bibinfo{year}{2004}{\natexlab{b}}), \eprint{astro-ph/0310725}.

\bibitem[{\citenamefont{McDonald et~al.}(2004)}]{McDonald:2004eu}
\bibinfo{author}{\bibfnamefont{P.}~\bibnamefont{McDonald}} \bibnamefont{et~al.}
  (\bibinfo{year}{2004}), \eprint{astro-ph/0405013}.

\bibitem[{\citenamefont{{McDonald} et~al.}(2005)\citenamefont{{McDonald},
  {Seljak}, {Cen}, {Shih}, {Weinberg}, {Burles}, {Schneider}, {Schlegel},
  {Bahcall}, {Briggs} et~al.}}]{McDonald:2004xn}
\bibinfo{author}{\bibfnamefont{P.}~\bibnamefont{{McDonald}}},
  \bibinfo{author}{\bibfnamefont{U.}~\bibnamefont{{Seljak}}},
  \bibinfo{author}{\bibfnamefont{R.}~\bibnamefont{{Cen}}},
  \bibinfo{author}{\bibfnamefont{D.}~\bibnamefont{{Shih}}},
  \bibinfo{author}{\bibfnamefont{D.~H.} \bibnamefont{{Weinberg}}},
  \bibinfo{author}{\bibfnamefont{S.}~\bibnamefont{{Burles}}},
  \bibinfo{author}{\bibfnamefont{D.~P.} \bibnamefont{{Schneider}}},
  \bibinfo{author}{\bibfnamefont{D.~J.} \bibnamefont{{Schlegel}}},
  \bibinfo{author}{\bibfnamefont{N.~A.} \bibnamefont{{Bahcall}}},
  \bibinfo{author}{\bibfnamefont{J.~W.} \bibnamefont{{Briggs}}},
  \bibnamefont{et~al.}, \bibinfo{journal}{\apj} \textbf{\bibinfo{volume}{635}},
  \bibinfo{pages}{761} (\bibinfo{year}{2005}).

\bibitem[{\citenamefont{Croft et~al.}(2002)}]{Croft:2000hs}
\bibinfo{author}{\bibfnamefont{R.~A.~C.} \bibnamefont{Croft}}
  \bibnamefont{et~al.}, \bibinfo{journal}{Astrophys. J.}
  \textbf{\bibinfo{volume}{581}}, \bibinfo{pages}{20} (\bibinfo{year}{2002}),
  \eprint{astro-ph/0012324}.

\bibitem[{\citenamefont{Viel et~al.}(2004{\natexlab{b}})\citenamefont{Viel,
  Haehnelt, and Springel}}]{Viel:2004bf}
\bibinfo{author}{\bibfnamefont{M.}~\bibnamefont{Viel}},
  \bibinfo{author}{\bibfnamefont{M.~G.} \bibnamefont{Haehnelt}},
  \bibnamefont{and} \bibinfo{author}{\bibfnamefont{V.}~\bibnamefont{Springel}},
  \bibinfo{journal}{Mon. Not. Roy. Astron. Soc.}
  \textbf{\bibinfo{volume}{354}}, \bibinfo{pages}{684}
  (\bibinfo{year}{2004}{\natexlab{b}}), \eprint{astro-ph/0404600}.

\bibitem[{\citenamefont{Hinshaw et~al.}(2003)}]{hinshaw03}
\bibinfo{author}{\bibfnamefont{G.}~\bibnamefont{Hinshaw}} \bibnamefont{et~al.},
  \bibinfo{journal}{\apjs} \textbf{\bibinfo{volume}{148}}, \bibinfo{pages}{135}
  (\bibinfo{year}{2003}), \eprint{astro-ph/0302217}.

\bibitem[{\citenamefont{Kogut et~al.}(2003)}]{kogut03}
\bibinfo{author}{\bibfnamefont{A.}~\bibnamefont{Kogut}} \bibnamefont{et~al.},
  \bibinfo{journal}{\apjs} \textbf{\bibinfo{volume}{148}}, \bibinfo{pages}{161}
  (\bibinfo{year}{2003}), \eprint{astro-ph/0302213}.

\bibitem[{\citenamefont{Readhead et~al.}(2004)}]{readhead04}
\bibinfo{author}{\bibfnamefont{A.~C.~S.} \bibnamefont{Readhead}}
  \bibnamefont{et~al.}, \bibinfo{journal}{\apj} \textbf{\bibinfo{volume}{609}},
  \bibinfo{pages}{498} (\bibinfo{year}{2004}), \eprint{astro-ph/0402359}.

\bibitem[{\citenamefont{Jones et~al.}(2005)}]{Jones:2005yb}
\bibinfo{author}{\bibfnamefont{W.~C.} \bibnamefont{Jones}} \bibnamefont{et~al.}
  (\bibinfo{year}{2005}), \eprint{astro-ph/0507494}.

\bibitem[{\citenamefont{Kuo et~al.}(2004)}]{Kuo:2002ua}
\bibinfo{author}{\bibfnamefont{C.-l.} \bibnamefont{Kuo}} \bibnamefont{et~al.}
  (\bibinfo{collaboration}{ACBAR}), \bibinfo{journal}{Astrophys. J.}
  \textbf{\bibinfo{volume}{600}}, \bibinfo{pages}{32} (\bibinfo{year}{2004}),
  \eprint{astro-ph/0212289}.

\bibitem[{\citenamefont{Dickinson et~al.}(2004)}]{dickinson04}
\bibinfo{author}{\bibfnamefont{C.}~\bibnamefont{Dickinson}}
  \bibnamefont{et~al.}, \bibinfo{journal}{\mnras}
  \textbf{\bibinfo{volume}{353}}, \bibinfo{pages}{732} (\bibinfo{year}{2004}),
  \eprint{astro-ph/0402498}.

\bibitem[{\citenamefont{Viel et~al.}(2005)\citenamefont{Viel, Lesgourgues,
  Haehnelt, Matarrese, and Riotto}}]{Viel:2005qj}
\bibinfo{author}{\bibfnamefont{M.}~\bibnamefont{Viel}},
  \bibinfo{author}{\bibfnamefont{J.}~\bibnamefont{Lesgourgues}},
  \bibinfo{author}{\bibfnamefont{M.~G.} \bibnamefont{Haehnelt}},
  \bibinfo{author}{\bibfnamefont{S.}~\bibnamefont{Matarrese}},
  \bibnamefont{and} \bibinfo{author}{\bibfnamefont{A.}~\bibnamefont{Riotto}},
  \bibinfo{journal}{Phys. Rev.} \textbf{\bibinfo{volume}{D71}},
  \bibinfo{pages}{063534} (\bibinfo{year}{2005}), \eprint{astro-ph/0501562}.

\bibitem[{\citenamefont{{Dalal} and {Kochanek}}(2002)}]{dalal2002}
\bibinfo{author}{\bibfnamefont{N.}~\bibnamefont{{Dalal}}} \bibnamefont{and}
  \bibinfo{author}{\bibfnamefont{C.~S.} \bibnamefont{{Kochanek}}},
  \bibinfo{journal}{\apj} \textbf{\bibinfo{volume}{572}}, \bibinfo{pages}{25}
  (\bibinfo{year}{2002}), \eprint{astro-ph/0111456}.

\bibitem[{\citenamefont{{Kochanek} and {Dalal}}(2004)}]{kochanek2004}
\bibinfo{author}{\bibfnamefont{C.~S.} \bibnamefont{{Kochanek}}}
  \bibnamefont{and} \bibinfo{author}{\bibfnamefont{N.}~\bibnamefont{{Dalal}}},
  \bibinfo{journal}{\apj} \textbf{\bibinfo{volume}{610}}, \bibinfo{pages}{69}
  (\bibinfo{year}{2004}), \eprint{astro-ph/0302036}.

\bibitem[{\citenamefont{Knebe et~al.}(2003)\citenamefont{Knebe, Devriendt,
  Gibson, and Silk}}]{Knebe:2003hs}
\bibinfo{author}{\bibfnamefont{A.}~\bibnamefont{Knebe}},
  \bibinfo{author}{\bibfnamefont{J.~E.~G.} \bibnamefont{Devriendt}},
  \bibinfo{author}{\bibfnamefont{B.~K.} \bibnamefont{Gibson}},
  \bibnamefont{and} \bibinfo{author}{\bibfnamefont{J.}~\bibnamefont{Silk}},
  \bibinfo{journal}{Mon. Not. Roy. Astron. Soc.}
  \textbf{\bibinfo{volume}{345}}, \bibinfo{pages}{1285} (\bibinfo{year}{2003}),
  \eprint{astro-ph/0302443}.

\bibitem[{\citenamefont{Chen et~al.}(2003)\citenamefont{Chen, Kravtsov, and
  Keeton}}]{Chen:2003uu}
\bibinfo{author}{\bibfnamefont{J.}~\bibnamefont{Chen}},
  \bibinfo{author}{\bibfnamefont{A.~V.} \bibnamefont{Kravtsov}},
  \bibnamefont{and} \bibinfo{author}{\bibfnamefont{C.~R.}
  \bibnamefont{Keeton}}, \bibinfo{journal}{Astrophys. J.}
  \textbf{\bibinfo{volume}{592}}, \bibinfo{pages}{24} (\bibinfo{year}{2003}),
  \eprint{astro-ph/0302005}.

\bibitem[{\citenamefont{{Barkana} et~al.}(2001)\citenamefont{{Barkana},
  {Haiman}, and {Ostriker}}}]{Barkana2001}
\bibinfo{author}{\bibfnamefont{R.}~\bibnamefont{{Barkana}}},
  \bibinfo{author}{\bibfnamefont{Z.}~\bibnamefont{{Haiman}}}, \bibnamefont{and}
  \bibinfo{author}{\bibfnamefont{J.~P.} \bibnamefont{{Ostriker}}},
  \bibinfo{journal}{Astrophys. J.} \textbf{\bibinfo{volume}{558}},
  \bibinfo{pages}{482} (\bibinfo{year}{2001}), \eprint{astro-ph/0102304}.

\bibitem[{\citenamefont{Yoshida et~al.}(2003)\citenamefont{Yoshida, Sokasian,
  Hernquist, and Springel}}]{Yoshida:2003rm}
\bibinfo{author}{\bibfnamefont{N.}~\bibnamefont{Yoshida}},
  \bibinfo{author}{\bibfnamefont{A.}~\bibnamefont{Sokasian}},
  \bibinfo{author}{\bibfnamefont{L.}~\bibnamefont{Hernquist}},
  \bibnamefont{and} \bibinfo{author}{\bibfnamefont{V.}~\bibnamefont{Springel}},
  \bibinfo{journal}{Astrophys. J.} \textbf{\bibinfo{volume}{591}},
  \bibinfo{pages}{L1} (\bibinfo{year}{2003}), \eprint{astro-ph/0303622}.

\bibitem[{\citenamefont{Rozas et~al.}(2005)\citenamefont{Rozas, Miralda-Escude,
  and Salvador-Sole}}]{Rozas:2005pt}
\bibinfo{author}{\bibfnamefont{J.~M.} \bibnamefont{Rozas}},
  \bibinfo{author}{\bibfnamefont{J.}~\bibnamefont{Miralda-Escude}},
  \bibnamefont{and}
  \bibinfo{author}{\bibfnamefont{E.}~\bibnamefont{Salvador-Sole}}
  (\bibinfo{year}{2005}), \eprint{astro-ph/0511489}.

\bibitem[{\citenamefont{Verde et~al.}(2003)}]{verde03}
\bibinfo{author}{\bibfnamefont{L.}~\bibnamefont{Verde}} \bibnamefont{et~al.},
  \bibinfo{journal}{\apjs} \textbf{\bibinfo{volume}{148}}, \bibinfo{pages}{195}
  (\bibinfo{year}{2003}), \eprint{astro-ph/0302218}.

\bibitem[{\citenamefont{Abazajian et~al.}(2001)\citenamefont{Abazajian, Fuller,
  and Tucker}}]{Abazajian:2001vt}
\bibinfo{author}{\bibfnamefont{K.}~\bibnamefont{Abazajian}},
  \bibinfo{author}{\bibfnamefont{G.~M.} \bibnamefont{Fuller}},
  \bibnamefont{and} \bibinfo{author}{\bibfnamefont{W.~H.}
  \bibnamefont{Tucker}}, \bibinfo{journal}{Astrophys. J.}
  \textbf{\bibinfo{volume}{562}}, \bibinfo{pages}{593} (\bibinfo{year}{2001}),
  \eprint{astro-ph/0106002}.

\bibitem[{\citenamefont{Lewis and Bridle}(2002)}]{lewis02}
\bibinfo{author}{\bibfnamefont{A.}~\bibnamefont{Lewis}} \bibnamefont{and}
  \bibinfo{author}{\bibfnamefont{S.}~\bibnamefont{Bridle}},
  \bibinfo{journal}{Phys. Rev.} \textbf{\bibinfo{volume}{D66}},
  \bibinfo{pages}{103511} (\bibinfo{year}{2002}), \eprint{astro-ph/0205436}.

\bibitem[{\citenamefont{{White} and {Rees}}(1978)}]{White1978}
\bibinfo{author}{\bibfnamefont{S.~D.~M.} \bibnamefont{{White}}}
  \bibnamefont{and} \bibinfo{author}{\bibfnamefont{M.~J.}
  \bibnamefont{{Rees}}}, \bibinfo{journal}{\mnras}
  \textbf{\bibinfo{volume}{183}}, \bibinfo{pages}{341} (\bibinfo{year}{1978}).

\bibitem[{\citenamefont{Weinberg et~al.}(2004)\citenamefont{Weinberg, Dave,
  Katz, and Hernquist}}]{Weinberg:2002rm}
\bibinfo{author}{\bibfnamefont{D.~H.} \bibnamefont{Weinberg}},
  \bibinfo{author}{\bibfnamefont{R.}~\bibnamefont{Dave}},
  \bibinfo{author}{\bibfnamefont{N.}~\bibnamefont{Katz}}, \bibnamefont{and}
  \bibinfo{author}{\bibfnamefont{L.}~\bibnamefont{Hernquist}},
  \bibinfo{journal}{Astrophys. J.} \textbf{\bibinfo{volume}{601}},
  \bibinfo{pages}{1} (\bibinfo{year}{2004}), \eprint{astro-ph/0212356}.

\bibitem[{\citenamefont{Zaldarriaga et~al.}(2003)\citenamefont{Zaldarriaga,
  Scoccimarro, and Hui}}]{Zaldarriaga:2001xs}
\bibinfo{author}{\bibfnamefont{M.}~\bibnamefont{Zaldarriaga}},
  \bibinfo{author}{\bibfnamefont{R.}~\bibnamefont{Scoccimarro}},
  \bibnamefont{and} \bibinfo{author}{\bibfnamefont{L.}~\bibnamefont{Hui}},
  \bibinfo{journal}{Astrophys. J.} \textbf{\bibinfo{volume}{590}},
  \bibinfo{pages}{1} (\bibinfo{year}{2003}), \eprint{astro-ph/0111230}.

\bibitem[{\citenamefont{Seljak et~al.}(2003)\citenamefont{Seljak, McDonald, and
  Makarov}}]{Seljak:2003jg}
\bibinfo{author}{\bibfnamefont{U.}~\bibnamefont{Seljak}},
  \bibinfo{author}{\bibfnamefont{P.}~\bibnamefont{McDonald}}, \bibnamefont{and}
  \bibinfo{author}{\bibfnamefont{A.}~\bibnamefont{Makarov}},
  \bibinfo{journal}{Mon. Not. Roy. Astron. Soc.}
  \textbf{\bibinfo{volume}{342}}, \bibinfo{pages}{L79} (\bibinfo{year}{2003}),
  \eprint{astro-ph/0302571}.

\bibitem[{\citenamefont{Jena et~al.}(2005)}]{Jena:2004fc}
\bibinfo{author}{\bibfnamefont{T.}~\bibnamefont{Jena}} \bibnamefont{et~al.},
  \bibinfo{journal}{Mon. Not. Roy. Astron. Soc.}
  \textbf{\bibinfo{volume}{361}}, \bibinfo{pages}{70} (\bibinfo{year}{2005}),
  \eprint{astro-ph/0412557}.

\bibitem[{\citenamefont{Viel and Haehnelt}(2006)}]{Viel:2005ha}
\bibinfo{author}{\bibfnamefont{M.}~\bibnamefont{Viel}} \bibnamefont{and}
  \bibinfo{author}{\bibfnamefont{M.~G.} \bibnamefont{Haehnelt}},
  \bibinfo{journal}{Mon. Not. Roy. Astron. Soc.}
  \textbf{\bibinfo{volume}{365}}, \bibinfo{pages}{231} (\bibinfo{year}{2006}),
  \eprint{astro-ph/0508177}.

\bibitem[{\citenamefont{Boyarsky et~al.}(2005)\citenamefont{Boyarsky, Neronov,
  Ruchayskiy, and Shaposhnikov}}]{Boyarsky:2005us}
\bibinfo{author}{\bibfnamefont{A.}~\bibnamefont{Boyarsky}},
  \bibinfo{author}{\bibfnamefont{A.}~\bibnamefont{Neronov}},
  \bibinfo{author}{\bibfnamefont{O.}~\bibnamefont{Ruchayskiy}},
  \bibnamefont{and}
  \bibinfo{author}{\bibfnamefont{M.}~\bibnamefont{Shaposhnikov}}
  (\bibinfo{year}{2005}), \eprint{astro-ph/0512509}.

\bibitem[{\citenamefont{{B{\"o}hringer} et~al.}(2001)}]{Boehringer2001}
\bibinfo{author}{\bibfnamefont{H.}~\bibnamefont{{B{\"o}hringer}}}
  \bibnamefont{et~al.}, \bibinfo{journal}{\aap} \textbf{\bibinfo{volume}{365}},
  \bibinfo{pages}{L181} (\bibinfo{year}{2001}), \eprint{astro-ph/0011459}.

\bibitem[{\citenamefont{Mapelli and Ferrara}(2005)}]{Mapelli:2005hq}
\bibinfo{author}{\bibfnamefont{M.}~\bibnamefont{Mapelli}} \bibnamefont{and}
  \bibinfo{author}{\bibfnamefont{A.}~\bibnamefont{Ferrara}}
  (\bibinfo{year}{2005}), \eprint{astro-ph/0508413}.

\end{thebibliography}

\end{document}